\documentclass[twocolumn,showpacs,preprintnumbers]{revtex4}
\usepackage{amssymb}
\usepackage{amsfonts}
\usepackage{float}
\usepackage{graphicx}
\usepackage{dcolumn}
\usepackage{bm}
\usepackage{epsfig}
\usepackage{amsmath}
\usepackage{subfigure}

\begin{document}

\title{Controllable single-photon nonreciprocal transmission in a cavity \\
optomechanical system with a weak coherent driving}
\author{Jun-Hao Liu}
\author{Ya-Fei Yu}
\email{yuyafei@m.scnu.edu.cn}
\author{Zhi-Ming Zhang}
\email{zhangzhiming@m.scnu.edu.cn}
\address{Guangdong Provincial Key Laboratory of Nanophotonic Functional Materials \& Devices (SIPSE), and
Guangdong Provincial Key Laboratory of Quantum Engineering \& Quantum Materials, South China
Normal University, Guangzhou 510006, China}
\begin{abstract}
We study the nonreciprocal transmission of a single-photon in a cavity
optomechanical system, in which the cavity supports a clockwise and a
counter-clockwise circulating optical modes, the mechanical resonator (MR)
is excited by a weak coherent driving, and the signal photon is made up of a
sequence of pulses with exactly one photon per pulse. We find that, if the
input state is a single-photon state, it is insufficient to study the
nonreciprocity only from the perspective of the transmission spectrums,
since the frequencies where the nonreciprocity happens are far away from the
peak frequency of the single-photon. So we show the nonreciprocal
transmission behavior by comparing the spectrums of the input and output
fields. In our system, we can achieve a transformation of the signal
transmission from unidirectional isolation to unidirectional amplification
in the single-photon level by changing the amplitude of the weak coherent
driving. The effects of the mechanical thermal noise on the single-photon
nonreciprocal transmission are also discussed.
\end{abstract}

\maketitle

\section{Introduction}

Nonreciprocal optical transmission has attracted more and more attention for
its important potential applications in quantum information processing and
quantum networks \cite{1}. In the nonreciprocal optical devices, e.g.,
isolator, circulator, nonreciprocal phase shifter, unidirectional amplifier,
the transmission of the information is not symmetric. Conventionally, the
nonreciprocal transmission can be achieved by using the Faraday rotation
effect in the magneto-optical crystals \cite{2,3,4,5,6,7}. However, this
scheme requires large magnetic fields, and thus make it difficult to realize
miniaturization and integration. In order to solve this problem, a number of
schemes have been proposed to break the reciprocity without the use of
magneto-optical effects. For example, one has proposed strategies that are
based on the optical nonlinearity \cite{8}, the spatial-symmetry-breaking
structures \cite{9,10,11,12}, the indirect interband photonic transitions
\cite{13,14,15,16,17,18,19}, the optoacoustic effects \cite{20,21}, the
parity-time-symmetric structures \cite{22,23,24,25}, and so on \cite{26,27}.

Recently, efforts have also been made to investigate the optical
nonreciprocity in cavity optomechanical systems \cite{28,29}. Manipatruni et
al. demonstrated that the optical nonreciprocal effect was based on the
momentum difference between the forward and backward-moving light beams in a
Fabry-Perot cavity with one moveable mirror \cite{30}. Subsequently, Hafezi
et al. proposed a scheme to achieve the nonreciprocal transmission in a
microring resonator by using a unidirectional optical pump \cite{31}. More
recently, many theoretical works aiming at achieving the circulator \cite{32}%
, the nonreciprocal quantum-state conversion \cite{33}, and the
unidirectional optical amplifier \cite{34} have been proposed in various of
cavity optomechanical systems. However, it is still a challenge to achieve
the nonreciprocal transmission in the single-photon or few-photon level. At
present, only a few works towards this target have been reported \cite%
{31,35,36}.

In this paper, we study the nonreciprocal transmission of a single-photon in
a cavity optomechanical system, as shown in Fig. 1, in which the cavity
supports a clockwise and a counter-clockwise circulating optical modes, the
mechanical resonator (MR) is excited by a weak coherent driving, and the
signal photon is made up of a sequence of pulses with exactly one photon per
pulse. We show that, it is insufficient to discuss the nonreciprocity from
the perspective of the transmission spectrums when we consider a
single-photon state as the input state, since the frequencies where the
nonreciprocity happens are far away from the peak frequency of the
single-photon. We will show the nonreciprocal transmission of the
single-photon by comparing the spectrums of the input and output fields. In
our system, we can achieve the unidirectional isolation and unidirectional
amplification of the signal in the single-photon level. Essentially, this
two kind of transmission behaviors are both caused by the unidirectional
optical pump, and the switch from the isolation to the amplification depends
on the amplitude of the weak coherent driving.

This paper is organized as follows. In Section II we introduce the
theoretical model, calculate the Langevin equations, derive the expressions
of the transmission spectrums and the spectrums of the output fields. In
Section III, we analyze the cause and condition of the unidirectional
isolation of the signal in the single-photon level by comparing the
spectrums of the input and output fields. Next in Section IV, we use the
same method to discuss the cause and condition of the unidirectional
amplification of the signal. In Section V, we consider the affect of the
mechanical noise on the nonreciprocal transmission, the experimental
feasibility of our system is also discussed in this section. Finally, we
provide a brief summary.

\begin{figure}[tbp]
%1%
\centering\includegraphics[width=8cm,height=4.98cm]{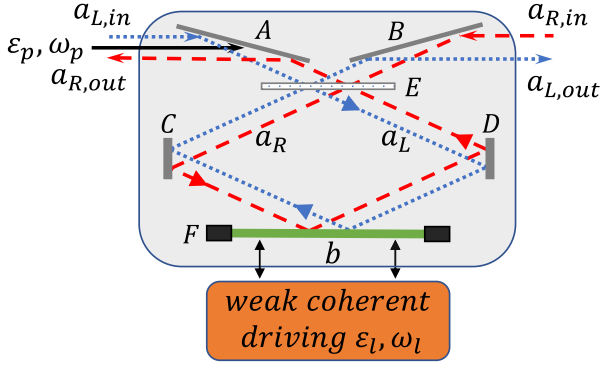}
\caption{(color online) Schematic diagram of our proposed model. A, B, C, D
are fixed mirrors of total reflection, E is a fixed mirrors of partial
transmission, and F is a movable mirror (treated as a mechanical resonator,
MR) of total reflection. C, D, E, and F constitute a ring resonator, which
supports a clockwise and a counter-clockwise circulating optical modes. The
MR is excited by a weak coherent driving. }
\end{figure}

\section{Model and Hamiltonian}

Our proposed scheme is shown in Fig. 1. A clockwise and a counter-clockwise
circulating cavity fields couple with the mechanical resonator via radiation
pressure. A strong coupling field $\varepsilon _{p}=\sqrt{2P\kappa /(\hbar
\omega _{p})}$ with frequency $\omega _{p}$ are injected from the left, in
which $P$ denotes its power. The MR is excited by a weak coherent driving
with amplitude $\varepsilon _{d}$ and frequency $\omega _{d}$, this driving
can be realized by, e.g., parametertically modulating the spring constant of
the MR at twice that MR's resonance frequency \cite{37,38,39,40}. The total
Hamiltonian of the system can be expressed as ($\hbar =1$)
\begin{eqnarray}
H_{total} &=&H_{coms}+H_{pump}+H_{driv}+H_{scat}, \\
H_{coms} &=&\omega _{c}\hat{a}_{L}^{\dag }\hat{a}_{L}+\omega _{c}\hat{a}%
_{R}^{\dag }\hat{a}_{R}+\omega _{m}\hat{b}^{\dag }\hat{b}  \notag \\
&&+g_{0}(\hat{a}_{L}^{\dag }\hat{a}_{L}+\hat{a}_{R}^{\dag }\hat{a}_{R})(\hat{%
b}^{\dag }+\hat{b}), \\
H_{pump} &=&i\varepsilon _{p}(\hat{a}_{L}^{\dag }e^{-i\omega _{p}t}-\hat{a}%
_{L}e^{i\omega _{p}t}), \\
H_{driv} &=&i\varepsilon _{d}[(\hat{b}^{\dag })^{2}e^{-i\omega _{d}t}-(\hat{b%
})^{2}e^{i\omega _{d}t}], \\
H_{scat} &=&J(\hat{a}_{L}^{\dag }\hat{a}_{R}+\hat{a}_{R}^{\dag }\hat{a}_{L}),
\end{eqnarray}%
where $H_{coms}$ describes the Hamiltonian of the cavity optomechanical
system, $\hat{a}_{L}(\hat{a}_{R})$ and $\hat{b}$ are the annihilation
operators of the clockwise (counter clockwise) circulating mode and the
mechanical mode with frequency $\omega _{c}$ and $\omega _{m}$,
respectively, $g_{0}$ is the optomechanical coupling strength between the
cavity field modes and the mechanical mode. $H_{pump}$ is the interaction
Hamiltonian between the strong coupling field and the clockwise circulating
mode. $H_{driv}$ is the interaction Hamiltonian between the weak coherent
driving and the MR. $H_{scat}$ represents a coherent scatting of strength $J$
between the two cavity modes, which is associated with the bulk or imperfect
reflection of the cavity.

In the rotation frame with $H$ $=$ $\omega _{p}(\hat{a}_{L}^{\dag }\hat{a}%
_{L}+\hat{a}_{R}^{\dag }\hat{a}_{R})$, we can obtain
\begin{eqnarray}
H_{T} &=&\Delta _{c}\hat{a}_{L}^{\dag }\hat{a}_{L}+\Delta _{c}\hat{a}%
_{R}^{\dag }\hat{a}_{R}+\omega _{m}\hat{b}^{\dag }\hat{b}+i\varepsilon _{p}(%
\hat{a}_{L}^{\dag }-\hat{a}_{L})  \notag \\
&&+g_{0}(\hat{a}_{L}^{\dag }\hat{a}_{L}+\hat{a}_{R}^{\dag }\hat{a}_{R})(\hat{%
b}^{\dag }+\hat{b})+J(\hat{a}_{L}^{\dag }\hat{a}_{R}+\hat{a}_{R}^{\dag }\hat{%
a}_{L})  \notag \\
&&+i\varepsilon _{d}[(\hat{b}^{\dag })^{2}e^{-i\omega _{d}t}-(\hat{b}%
)^{2}e^{i\omega _{d}t}],
\end{eqnarray}%
where $\Delta _{c}=\omega _{c}-\omega _{p}$ is the frequency detuning
between the cavity field and the coupling field. The system dynamics is
fully described by the set of quantum Langevin equations
\begin{eqnarray}
\frac{d\hat{a}_{L}}{dt} &=&-(i\Delta _{c}+\kappa _{t})\hat{a}_{L}-ig_{0}\hat{%
a}_{L}(\hat{b}^{\dag }+\hat{b})  \notag \\
&&-iJ\hat{a}_{R}+\varepsilon _{p}+\sqrt{2\kappa }\hat{a}_{L,in}, \\
\frac{d\hat{a}_{R}}{dt} &=&-(i\Delta _{c}+\kappa _{t})\hat{a}_{R}-ig_{0}\hat{%
a}_{R}(\hat{b}^{\dag }+\hat{b})  \notag \\
&&-iJ\hat{a}_{L}+\sqrt{2\kappa }\hat{a}_{R,in}, \\
\frac{d\hat{b}}{dt} &=&-(\gamma +i\omega _{m})\hat{b}-ig_{0}(\hat{a}%
_{L}^{\dag }\hat{a}_{L}+\hat{a}_{R}^{\dag }\hat{a}_{R})  \notag \\
&&+2\varepsilon _{d}e^{-i\omega _{d}t}\hat{b}^{\dag }+\sqrt{2\gamma }\hat{b}%
_{in},
\end{eqnarray}%
where the cavity has the damping rate $\kappa _{t}=\kappa _{in}+\kappa $,
which are assumed to be due to the intrinsic photon loss and external
coupling dissipation, respectively. The mechanical mode has the damping rate
$\gamma $ with the mechanical input operator $\hat{b}_{in}$ satisfying $%
\left\langle \hat{b}_{in}^{\dag }(\Omega )\hat{b}_{in}(\omega )\right\rangle
$ $=$ $n_{th}\delta (\Omega +\omega )$, $\left\langle \hat{b}_{in}(\Omega )%
\hat{b}_{in}^{\dag }(\omega )\right\rangle =(n_{th}+1)\delta (\Omega +\omega
)$ in the frequency domain, where $n_{th}$ is the thermal phonon occupation
number at a finite temperature. $\hat{a}_{L,in}$ and $\hat{a}_{R,in}$ are
the operators of the input fields from the left and right, respectively. We
consider the case in which the input field is made up of a sequence of
pulses with exactly one photon per pulse. The input field is centered near
the cavity frequency with a finite bandwidth, its spectrum is given by \cite%
{41,42} $S_{k,in}(\omega )$ $=$ $\frac{\Gamma /\pi }{(\omega -\omega
_{c})^{2}+\Gamma ^{2}}$ ($k$ $=$ $R$, $L$), and we assume that the decay
rate of the single-photon $\Gamma =\kappa $. The correlation functions of
the input operators in the frequency domain are (see the appendix)
\begin{align}
\left\langle \hat{a}_{k,in}^{\dag }(\Omega )\hat{a}_{k,in}(\omega
)\right\rangle & =S_{k,in}(\omega )\delta (\omega +\Omega ), \\
\left\langle \hat{a}_{k,in}(\Omega )\hat{a}_{k,in}^{\dag }(\omega
)\right\rangle & =[S_{k,in}(\Omega )+1]\delta (\omega +\Omega ).
\end{align}

Equations (7)-(9) can be solved by using the perturbation method in the
limit of a strong coupling field $\varepsilon _{p}$, while taking the
driving field $\varepsilon _{d}$ to be weak. We make a transformations $\hat{%
a}_{k}\rightarrow \hat{a}_{k}+\alpha _{k}$ ($k=L,R$), and $\hat{b}%
\rightarrow \hat{b}+\beta $ for all the interaction modes, then we can
obtain the steady state value equations
\begin{eqnarray}
0 &=&-(\gamma +i\omega _{m})\beta -ig_{0}(\left\vert \alpha _{L}\right\vert
^{2}+\left\vert \alpha _{R}\right\vert ^{2}), \\
0 &=&-(i\Delta _{c}+\kappa _{t})\alpha _{R}-ig_{0}\alpha _{R}(\beta +\beta
^{\ast })-iJ\alpha _{L}, \\
0 &=&-(i\Delta _{c}+\kappa _{t})\alpha _{L}-ig_{0}\alpha _{L}(\beta +\beta
^{\ast })  \notag \\
&&-iJ\alpha _{R}+\varepsilon _{p}.
\end{eqnarray}%
We assume that the system works near the red sideband ($\Delta _{c}$ $=$ $%
\omega _{m}$), since the optomechanical coupling strength $g_{0}$ and the
coherent scatting strength $J$ are both very weak. For a strong coupling
field $\varepsilon _{p}$, we can obtain $\alpha _{L}\gg \alpha _{R}$ and $%
\beta \simeq 0$ from the steady state equations.

Next, we can write out the linearized Hamiltonian of the system
\begin{eqnarray}
H_{l} &=&\Delta ^{\prime }(\hat{a}_{L}^{\dag }\hat{a}_{L}+\hat{a}_{R}^{\dag }%
\hat{a}_{R})+\omega _{m}\hat{b}^{\dag }\hat{b}+J(\hat{a}_{L}^{\dag }\hat{a}%
_{R}+\hat{a}_{R}^{\dag }\hat{a}_{L})  \notag \\
&&+g_{0}(\hat{b}^{\dag }+\hat{b})(\hat{a}_{L}^{\dag }\alpha _{L}+\hat{a}%
_{L}\alpha _{L}^{\ast }+\hat{a}_{R}^{\dag }\alpha _{R}+\hat{a}_{R}\alpha
_{R}^{\ast })  \notag \\
&&+i\varepsilon _{d}[(\hat{b}^{\dag })^{2}e^{-i\omega _{d}t}-(\hat{b}%
)^{2}e^{i\omega _{d}t}],
\end{eqnarray}%
where $\Delta ^{\prime }=\Delta _{c}+g_{0}(\beta +\beta ^{\ast })$. Because $%
\Delta _{c}\gg g_{0}(\beta +\beta ^{\ast }),$ we will assume that $\Delta
^{\prime }\approx \Delta _{c}$ in the following calculation. In the rotation
frame with $H^{\prime }=\frac{\omega _{d}}{2}(\hat{a}_{L}^{\dag }\hat{a}_{L}+%
\hat{a}_{R}^{\dag }\hat{a}_{R}+\hat{b}^{\dag }\hat{b})$, we have
\begin{eqnarray}
H_{l}^{\prime } &=&\delta _{c}(\hat{a}_{L}^{\dag }\hat{a}_{L}+\hat{a}%
_{R}^{\dag }\hat{a}_{R})+g_{0}(\alpha _{R}\hat{a}_{R}^{\dag }\hat{b}+\alpha
_{R}^{\ast }\hat{a}_{R}\hat{b}^{\dag })  \notag \\
&&+\Delta _{m}\hat{b}^{\dag }\hat{b}+g_{0}(\alpha _{L}\hat{a}_{L}^{\dag }%
\hat{b}+\alpha _{L}^{\ast }\hat{a}_{L}\hat{b}^{\dag })  \notag \\
&&+i\varepsilon _{d}[(\hat{b}^{\dag })^{2}-(\hat{b})^{2}]+J(\hat{a}%
_{L}^{\dag }\hat{a}_{R}+\hat{a}_{R}^{\dag }\hat{a}_{L}),
\end{eqnarray}%
where $\delta _{c}=\Delta _{c}-\omega _{d}/2,$ $\Delta _{m}=\omega
_{m}-\omega _{d}/2$. In addition, using the rotating-wave approximation, we
have omitted the high-frequency oscillation terms, such as $\hat{a}%
_{R}^{\dag }\hat{b}^{\dag }e^{i\omega _{d}t}$, $\hat{a}_{L}^{\dag }\hat{b}%
^{\dag }e^{i\omega _{d}t}$ and so on.

We define a vector $v(t)$ $=$ ($\hat{a}_{L}(t)$, $\hat{a}_{R}(t)$, $\hat{b}%
(t)$, $\hat{a}_{L}^{\dag }(t)$, $\hat{a}_{R}^{\dag }(t)$, $\hat{b}^{\dag
}(t) $$)^{T}$ in terms of the operators of the system modes. By substituting
$v(t) $ and $H_{l}^{\prime }$ into the quantum Langevin equation, we can
obtain
\begin{equation}
\frac{dv(t)}{dt}=Mv(t)+\sqrt{2\kappa }v_{L,in}+\sqrt{2\kappa }v_{R,in}+\sqrt{%
2\gamma }v_{b,in},
\end{equation}%
where $v_{L,in}$ $=$ $($$\hat{a}_{L,in}(t)$, $0$, $0$, $\hat{a}_{L,in}^{\dag
}(t)$, $0$, $0)^{T}$, $v_{R,in}$ $=$ $($$0$, $\hat{a}_{R,in}(t)$, $0$, $0$, $%
\hat{a}_{R,in}^{\dag }(t)$, $0)^{T}$, $v_{b,in}$$=$ $($$0$, $0$, $\hat{b}%
_{in}(t)$, $0$, $0$, $\hat{b}_{in}^{\dag }(t))^{T}$, and
\begin{equation}
M=\left(
\begin{array}{cccccc}
-\phi & -iJ & -iG_{L} & 0 & 0 & 0 \\
-iJ & -\phi & -iG_{R} & 0 & 0 & 0 \\
-iG_{L}^{\ast } & -iG_{R}^{\ast } & -\varphi & 0 & 0 & 2\varepsilon _{d} \\
0 & 0 & 0 & \phi & iJ & iG_{L}^{\ast } \\
0 & 0 & 0 & iJ & \phi & iG_{R}^{\ast } \\
0 & 0 & 2\varepsilon _{d} & iG_{L} & iG_{R} & \varphi%
\end{array}%
\right) ,
\end{equation}%
where $\pm \phi $ $=$ $-\kappa _{t}\pm i\delta _{c}$, $\pm \varphi $ $=$ $%
-\gamma \pm i\Delta _{m}$, $G_{k}=g_{0}\alpha _{k}$ ($k=R,L$). Without loss
of generality, we take $G_{k}$ as a real number in the following
calculation. The system is stable only if the real parts of all the
eigenvalues of matrix $M$ are negative. The stability conditions can be
explicitly given by using the Routh-Hurwitz criterion \cite{43,44,45}.
However, they are too verbose to be given here, and we make sure the
stability conditions are fulfilled in the system with our used parameters.

By introducing the Fourier transform of the operators
\begin{eqnarray}
\hat{o}(\omega ) &=&\int_{-\infty }^{+\infty }\hat{o}(t)e^{i\omega t}dt, \\
\hat{o}^{\dag }(\omega ) &=&\int_{-\infty }^{+\infty }\hat{o}^{\dag
}(t)e^{i\omega t}dt,
\end{eqnarray}%
where $o=a_{L},a_{R},b$, we can solve the linearized quantum Langevin
equations (17) in the frequency domain
\begin{eqnarray}
v(\omega ) &=&-(M+i\omega I)^{-1}[\sqrt{2\kappa }v_{L,in}(\omega )  \notag \\
&&+\sqrt{2\kappa }v_{R,in}(\omega )+\sqrt{2\gamma }v_{b,in}(\omega )],
\end{eqnarray}%
where $v(\omega )$ $=$ $($$\hat{a}_{L}(\omega )$, $\hat{a}_{R}(\omega )$, $%
\hat{b}(\omega )$, $\hat{a}_{L}^{\dag }(\omega )$, $\hat{a}_{R}^{\dag
}(\omega )$, $\hat{b}^{\dag }(\omega ))^{T}$, $v_{L,in}(\omega )$ $=$ $(\hat{%
a}_{L,in}(\omega )$, $0$, $0$, $\hat{a}_{L,in}^{\dag }(\omega )$, $0$, $%
0)^{T}$, $v_{R,in}(\omega )$ $=$ $($$0$, $\hat{a}_{R,in}(\omega )$, $0$, $0$%
, $\hat{a}_{R,in}^{\dag }(\omega )$, $0)^{T}$, and $v_{b,in}(\omega )$ $=$ $%
( $$0$, $0$, $\hat{b}_{in}(\omega )$, $0$, $0$, $\hat{b}_{in}^{\dag }(\omega
))^{T}$. As a consequence of boundary conditions, the relation among the
input, internal, and output fields is given as \cite{46}
\begin{equation}
\hat{a}_{k,out}(\omega )=-\hat{a}_{k,in}(\omega )+\sqrt{2\kappa }\hat{a}%
_{k}(\omega ),k=R,L.
\end{equation}%
From Eq. (21) and Eq. (22), we can write the operators of the output fields
as
\begin{equation}
\hat{a}_{L,out}(\omega )=\mathit{f}^{L}(\omega )v_{in}(\omega ),\hat{a}%
_{R,out}(\omega )=\mathit{f}^{R}(\omega )v_{in}(\omega ),
\end{equation}%
where $\mathit{f}^{k}(\omega )$ $=$ $($$f_{1}^{k}(\omega )$, $%
f_{2}^{k}(\omega )$, $f_{3}^{k}(\omega )$, $f_{4}^{k}(\omega )$, $%
f_{5}^{k}(\omega )$, $f_{6}^{k}(\omega ))$ ($k$ $=$ $R$, $L$), $%
v_{in}(\omega )$ $=$ $($$\hat{a}_{L,in}(\omega )$, $\hat{a}_{R,in}(\omega )$%
, $\hat{b}_{in}(\omega )$, $\hat{a}_{L,in}^{\dag }(\omega )$, $\hat{a}%
_{R,in}^{\dag }(\omega )$, $\hat{b}_{in}^{\dag }(\omega ))^{T}$, in which
the concrete form of the coefficients $\mathit{f}^{L}(\omega )$ and $\mathit{%
f}^{R}(\omega )$ are tediously long, we will not write out here.

The spectrums of the output fields are defined by
\begin{equation}
S_{k,out}(\omega )=\int d\omega \left\langle \hat{a}_{k,out}^{\dag }(\Omega )%
\hat{a}_{k,out}(\omega )\right\rangle ,k=R,L.
\end{equation}%
By substituting the expressions of $\hat{a}_{R,out}(\omega )$ and $\hat{a}%
_{L,out}(\omega )$ into Eq. (24), and using the correlation functions, one
can obtain
\begin{eqnarray}
S_{L,out}(\omega ) &=&F_{1}^{L}S_{L,in}(\omega )+F_{2}^{L}S_{R,in}(\omega
)+F_{3}^{L}n_{th}  \notag \\
&&+F_{4}^{L}S_{L,vac}(-\omega )+F_{5}^{L}S_{R,vac}(-\omega )  \notag \\
&&+F_{6}^{L}(n_{th}+1). \\
S_{R,out}(\omega ) &=&F_{1}^{R}S_{L,in}(\omega )+F_{2}^{R}S_{R,in}(\omega
)+F_{3}^{R}n_{th}  \notag \\
&&+F_{4}^{R}S_{L,vac}(-\omega )+F_{5}^{R}S_{R,vac}(-\omega )  \notag \\
&&+F_{6}^{R}(n_{th}+1),
\end{eqnarray}%
where $F_{j}^{k}=\left\vert f_{j}^{k}(\omega )\right\vert ^{2}$, $%
S_{k,vac}(-\omega )=S_{k,in}(-\omega )+1$ ($k=R,L$, $j=1,2,3,4,5,6$). We can
see that the spectrum of the output fields $S_{L,out}(\omega )$ and $%
S_{R,out}(\omega )$ both contain six components. For $S_{R,out}(\omega )$, $%
F_{1}^{R}$ and $F_{2}^{R}$ represent the scattering probability of the input
fields $S_{L,in}(\omega )$ and $S_{R,in}(\omega )$, respectively. $F_{3}^{R}$
is the scattering probability of the mechanical thermal noise. $F_{4}^{R}$, $%
F_{5}^{R}$, and $F_{6}^{R}$ denote the scattering probability of the vacuum
fluctuations of their corresponding input fields.

In this paper, the parameters used are $\omega _{m}$ $=$ $25$ MHz and $%
\gamma $ $=$ $100$ Hz (quality factor $Q_{m}$ $=$ $2.5\times 10^{5}$). The
damping rate of the optical cavity $\kappa $ $=$ $1$ MHz, $g_{0}$ $=$ $1$
kHz, and the enhanced optomechanical coupling strength $G_{L}$ $=$ $16$ MHz.
The other parameters are $G_{R}$ $=$ $1$ kHz, $J$ $=$ $10$ kHz, $\kappa
_{in} $ $=$ $1$ MHz.

\begin{figure}[b]
%1%
\centering\includegraphics[width=8cm,height=5.5cm]{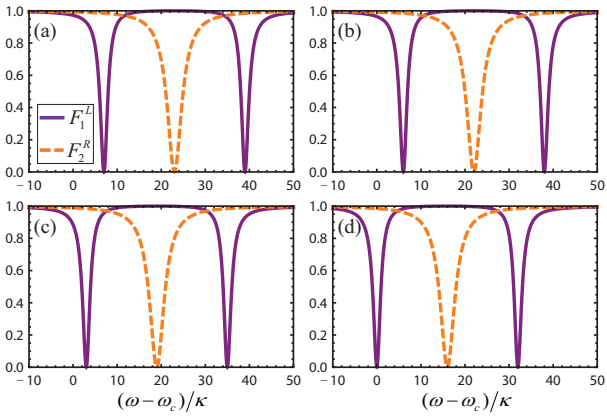}
\caption{(color online) The spectrums of the scattering probabilities $%
F_{1}^{L}$ (purple solid lines) and $F_{2}^{R}$ (orange dashed lines) for
driving frequency $\protect\omega _{d}$: (a) $\protect\omega _{d}/\protect%
\kappa $ $=$ $4$, (b) $\protect\omega _{d}/\protect\kappa $ $=$ $6$, (c) $%
\protect\omega _{d}/\protect\kappa $ $=$ $12$, (d) $\protect\omega _{d}/%
\protect\kappa $ $=$ $18$. The other parameters are stated in the text. }
\end{figure}

\begin{figure}[tph]
%1%
\centering\includegraphics[width=8cm,height=11.68cm]{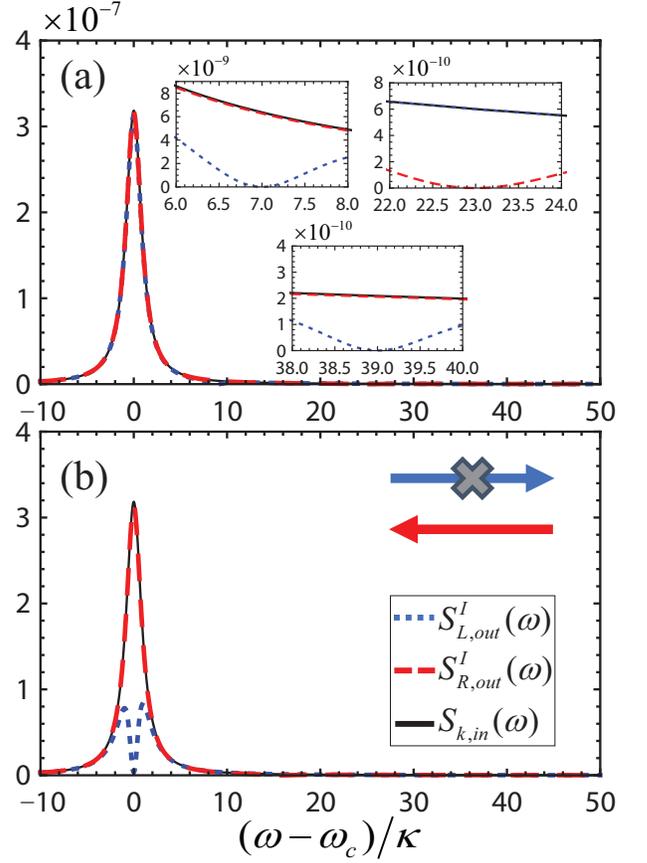}
\caption{(color online) The spectrums of the output fields $S_{L,out}^{I}(%
\protect\omega )$ (blue dotted lines), $S_{R,out}^{I}(\protect\omega )$ (red
dashed lines) and the input fields $S_{k,in}(\protect\omega )$ (Black solid
lines) for different driving frequency $\protect\omega _{d}$: (a) $\protect%
\omega _{d}/\protect\kappa $ $=$ $4$, (b) $\protect\omega _{d}/\protect%
\kappa =18$. The other parameters are given in the text.}
\end{figure}

\section{Unidirectional isolation of the signal in the single-photon level}

In this section, we numerically evaluate the scattering probabilities and
the spectrums of the input-output fields to show the possibility of
achieving the unidirectional isolation of the signal in the single-photon
level. It should be pointed out that, we have plotted the spectrums of all
the scattering probabilities $F_{1}^{L},\cdots ,F_{6}^{L}$ and $%
F_{1}^{R},\cdots ,F_{6}^{R}$, and found that in the range of the parameters
we considered ($\varepsilon _{d}/\kappa $ $=$ $6$ $\times $ $10^{-5}$), the
scattering probabilities have following order of magnitude: $F_{1}^{L}$ $%
\sim $ $1$, $F_{4}^{L}$ $\sim $ $10^{-9}$, $F_{2}^{L}$ $\sim $ $10^{-7}$, $%
F_{5}^{L}$ $\sim $ $10^{-15}$, $F_{3}^{L}$ $\sim $ $10^{-4}$, $F_{6}^{L}$ $%
\sim $ $10^{-13}$, and $F_{1}^{R}$ $\sim $ $10^{-7}$, $F_{4}^{R}$ $\sim $ $%
10^{-15}$, $F_{2}^{R}$ $\sim $ $1$, $F_{5}^{R}$ $\sim $ $10^{-22}$, $%
F_{3}^{R}$ $\sim $ $10^{-11}$, $F_{6}^{R}$ $\sim $ $10^{-19}$. In the
single-photon level, the peak value of the spectrum of the input field $%
S_{k,in}(\omega )$ $\sim $ $10^{-7}$. Hence the spectrums of the output
fields can be reduced to
\begin{equation}
S_{L,out}^{I}(\omega )=F_{1}^{L}S_{L,in}(\omega ),S_{R,out}^{I}(\omega
)=F_{2}^{R}S_{R,in}(\omega ),
\end{equation}%
and we have assumed that the thermal phonon occupation number $n_{th}=0$.

As shown in Fig. 2(a), we plot the spectrums of the scattering probabilities
$F_{1}^{L}$ and $F_{2}^{R}$ for different driving frequency $\omega _{d}$.
We can see that the transmission of the left-going mode is simply that of a
bare resonator, while the transmission of the right-going mode is modified
by the presence of the MR, the effective optomechanical coupling $G_{L}$
will lead to a normal mode splitting \cite{47,48} in the strong coupling
regime. With the presence of the weak coherent driving, the effective
frequency of the mechanical mode becomes $\Delta _{m}=\omega _{m}-\omega
_{d}/2.$ The above features will result in a unidirectional isolation
between the left-going mode and right-going mode at three positions: at $%
\omega -\omega _{c}$ $=$ $\Delta _{m}$, we have $F_{1}^{L}$ $=$ $1$, $%
F_{2}^{R}$ $=$ $0$; at $\omega -\omega _{c}$ $=$ $\Delta _{m}$ $\pm $ $G_{L}$%
, we have $F_{1}^{L}$ $=$ $0$, $F_{2}^{R}$ $\lesssim $ $1$. When we increase
$\omega _{d}$, the effective frequency $\Delta _{m}$ will decrease, and the curves will integrally move to the left.

To see the nonreciprocal transmission from the perspective of the spectrums
of the output fields, we should also consider the spectrum of the input
field. In Fig. 3, we plot the spectrum of the input field $S_{k,in}(\omega )$
as a comparison with the spectrums of the output fields. In Fig. 3(a), we
use the same parameters as that in Fig. 2(a). It can be seen that, the
non-reciprocity will indeed emerge at $\omega -\omega _{c}$ $=$ $\Delta _{m}$
and $\omega -\omega _{c}$ $=$ $\Delta _{m}\pm $ $G_{L}$. However, these
points are all far away from the peak frequency of the input field, the
values of $S_{k,in}(\omega )$ at these points are too small. The spectrums
of the output fields $S_{L,out}^{I}(\omega )$, $S_{R,out}^{I}(\omega )$ and
the input field $S_{k,in}(\omega )$ are almost coincide, and the
nonreciprocity in this case can be ignored. By adjusting the driving
frequency $\omega _{d}$, we can move the frequency where the non-reciprocity
happens to the peak frequency of the single-photon. As shown in Fig. 3(b),
we use the same parameters as that in Fig. 2(d). It can be seen that, at $%
\omega -\omega _{c}$ $=$ $0$, $S_{L,out}^{I}(\omega )$ $=0$, $%
S_{R,out}^{I}(\omega )$ $=$ $S_{k,in}(\omega )$. In this case, the
nonreciprocity is very obvious. The signal can transmit from the right to
the left, but can not transmit in the opposite direction. In this case, our
system can act as a unidirectional isolator in the single-photon level.

\begin{figure}[t]
%1%
\centering\includegraphics[width=8cm,height=5.85cm]{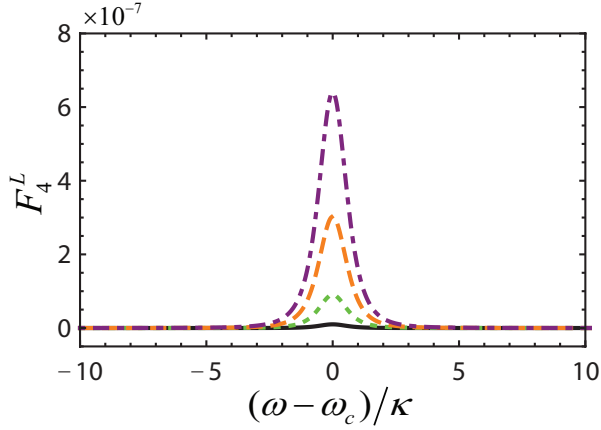}
\caption{(color online) The spectrums of the scattering probabilities $%
F_{4}^{L}$ for different driving amplitude $\protect\varepsilon _{d}$: (a) $%
\protect\varepsilon _{d}/\protect\kappa =1\times 10^{-4}$ (black solid
line), (b) $\protect\varepsilon _{d}/\protect\kappa =3\times 10^{-4}$ (green
dotted line), (c) $\protect\varepsilon _{d}/\protect\kappa =5.5\times
10^{-4} $ (orange dashed line), (d) $\protect\varepsilon _{d}/\protect\kappa %
=8\times 10^{-4}$ (purple dotdashed line), the other parameters are the same
as in Fig. 2(d).}
\end{figure}

\begin{figure}[tph]
%1%
\centering\includegraphics[width=8cm,height=16.92cm]{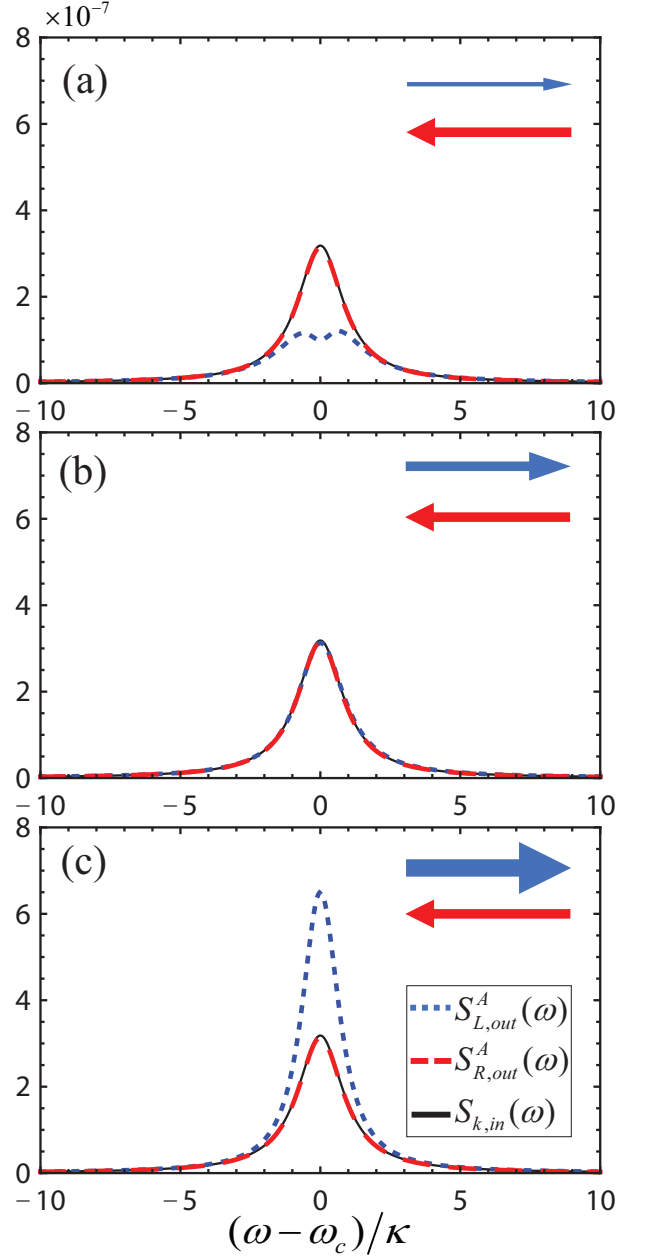}
\caption{(color online) The spectrums of the output fields $S_{L,out}^{A}(%
\protect\omega )$ (blue dotted lines), $S_{R,out}^{A}(\protect\omega )$ (red
dashed lines) and the input fields $S_{k,in}(\protect\omega )$ (Black solid
lines) for different driving amplitude $\protect\varepsilon _{d}$ (a) $%
\protect\varepsilon _{d}/\protect\kappa =3\times 10^{-4}$, (b) $\protect%
\varepsilon _{d}/\protect\kappa =5.5\times 10^{-4}$, (c) $\protect%
\varepsilon _{d}/\protect\kappa =8\times 10^{-4}$. The other parameters are
the same as that in Fig. 4.}
\end{figure}

\section{Unidirectional amplification of the signal in the single-photon
level}

A previous work \cite{40} has suggested that, such a weak coherent driving
can induce a remarkable enhancement of the output fields. In this section,
we will numerically evaluate the scattering probabilities and the spectrums
of the input and output fields to show the possibility of achieving the
unidirectional amplification of the signal in the single-photon level.
Likewise, we have also plotted the spectrums of all the scattering
probabilities $F_{1}^{L},\cdots ,F_{6}^{L}$ and $F_{1}^{R},\cdots ,F_{6}^{R}$%
. We find that, in the range of the parameters we considered ($1$ $\times $ $%
10^{-4}$ $\leq $ $\varepsilon _{d}/\kappa $ $\leq $ $8$ $\times $ $10^{-4}$%
), the scattering probabilities have the order of magnitude: $F_{1}^{L}$ $%
\sim $ $1$, $F_{4}^{L}$ $\sim $ $10^{-7}$, $F_{2}^{L}$ $\sim $ $10^{-7}$, $%
F_{5}^{L}$ $\sim $ $10^{-13}$, $F_{3}^{L}$ $\sim $ $10^{-4}$, $F_{6}^{L}$ $%
\sim $ $10^{-11}$, and $F_{1}^{R}$ $\sim $ $10^{-7}$, $F_{4}^{R}$ $\sim $ $%
10^{-13}$, $F_{2}^{R}$ $\sim $ $1$, $F_{5}^{R}$ $\sim $ $10^{-20}$, $%
F_{3}^{R}$ $\sim $ $10^{-11}$, $F_{6}^{R}$ $\sim $ $10^{-17}$. The peak
value of the spectrum of the input field $S_{k,in}(\omega )$ $\sim $ $10^{-7}
$. Hence the spectrums of the output fields can be reduced to
\begin{eqnarray}
S_{L,out}^{A}(\omega ) &=&F_{1}^{L}S_{L,in}(\omega )+F_{4}^{L}, \\
S_{R,out}^{A}(\omega ) &=&F_{2}^{R}S_{R,in}(\omega ),
\end{eqnarray}%
in which we have assumed that the thermal phonon occupation number $n_{th}=0$%
.

Since the amplitude of the weak coherent driving is very weak, we find that
the scattering probabilities $F_{1}^{L}$ and $F_{2}^{R}$ are almost
unchanged with the increase of $\varepsilon _{d}$. However, the weak
coherent driving will induced a remarkable amplification on the scattering
probabilities $F_{4}^{L}$, as shown in Fig. 4, with the increase of $%
\varepsilon _{d}$, at $\omega -\omega _{c}$ $=$ $0$, $F_{4}^{L}$ will
gradually increase. This feature can be used to achieve the unidirectional
amplification of the signal in the single-photon level.

From the perspective of the spectrums of the output fields, as shown in Fig.
5, with the increase of the driving amplitude $\varepsilon _{d}$, the value
of the spectrum of the output field $S_{L,out}^{A}(\omega )$ at $\omega
-\omega _{c}$ $=$ $0$ will increase, and the nonreciprocity will be weaken.
When $\varepsilon _{d}$ reaches a certain threshold ($\varepsilon
_{d}/\kappa $ $=$ $5.5\times 10^{-4}$), the nonreciprocity will almost
disappear, i.e., $S_{L,out}^{A}(\omega )$ $\approx $ $S_{R,out}^{A}(\omega )$
$\approx $ $S_{k,in}(\omega )$. Furthermore, if we continue to increase the
driving amplitude $\varepsilon _{d}$, the system will reveal a nonreciprocal
amplification phenomenon as shown in Fig. 5(c), the signal transmitted from
left to right can be amplified ($S_{R,out}^{A}(\omega )$ $>$ $%
S_{k,in}(\omega )$ at $\omega -\omega _{c}$ $=$ $0$), while the signal
transmitted from right to left cannot be amplified ($S_{L,out}^{A}(\omega )$
$\approx $ $S_{k,in}(\omega )$ at $\omega -\omega _{c}$ $=$ $0$). In this
case, our system can act as a unidirectional amplifier in the single-photon
level.

\section{Discussion and Conclusion}

Now we consider the effects of the mechanical thermal noise on the
unidirectional isolator and amplifier. When the thermal phonon occupation
number $n_{th}\neq 0$, the spectrums of the output fields become
\begin{eqnarray}
S_{L,out}^{I}(\omega ) &=&F_{1}^{L}S_{L,in}(\omega )+F_{3}^{L}n_{th}, \\
S_{R,out}^{I}(\omega ) &=&F_{2}^{R}S_{R,in}(\omega )+F_{3}^{R}n_{th}, \\
S_{L,out}^{A}(\omega ) &=&F_{1}^{L}S_{L,in}(\omega
)+F_{4}^{L}+F_{3}^{L}n_{th}, \\
S_{R,out}^{A}(\omega ) &=&F_{2}^{R}S_{R,in}(\omega )+F_{3}^{R}n_{th}.
\end{eqnarray}%
If the influence of the mechanical thermal noise can be neglected, we should
ensure $F_{3}^{L}n_{th}$ $\ll $ $F_{1}^{L}S_{L,in}(\omega )$ and $%
F_{3}^{R}n_{th}$ $\ll $ $F_{2}^{R}S_{R,in}(\omega )$. From the above
discussion, we have $F_{1}^{L}$, $F_{2}^{R}$ $\sim $ $1$, $F_{3}^{L}$ $\sim $
$10^{-4}$, $F_{3}^{R}$ $\sim $ $10^{-11}$, and $S_{k,in}(\omega )$ $\sim $ $%
10^{-7}$. Hence we should guarantee the thermal phonon occupation number $%
n_{th}$ $<$ $n_{thres}$ $\sim $ $10^{-4}$, i.e., the mechanical resonator
should be cooled near its quantum ground state. However, if we choose a
single-photon whose spectrum is narrower than the linewidth of the cavity,
the threshold $n_{thres}$ can be improved. For example, if we choose $\Gamma
=0.005\kappa ,$ $S_{k,in}(\omega )$ can reach the order of magnitude $%
10^{-5} $, and $n_{thres}$ will be increased to $10^{-2}$. In addition, we
can also increase the threshold $n_{thres}$ by improving the quality factor
of the MR, when $Q_{m}$ $\sim $ $10^{8}$, $F_{3}^{L}$ will be reduced to $%
10^{-7}$, if $S_{k,in}(\omega )$ $\sim $ $10^{-7}$, $n_{thres}$ can reach
the order of magnitude $10^{-1}$.

In summary, we have studied the single-photon nonreciprocal transmission in
a cavity optomechanical system, in which the mechanical resonator is
exciting by a weak coherent driving. We have shown that, if the input state
is a single-photon state, it is insufficient to study the nonreciprocity
only from the perspective of the transmission spectrums. Our scheme can be
used as a unidirectional isolator or amplifier in the single-photon level.
Our proposed model might eventually provide the basis for the applications
on quantum information processing or quantum networks.

\section*{ACKNOWLEDGEMENTS}

\addcontentsline{toc}{section}{Acknowledgements} This work was supported by
the National Natural Science Foundation of China (Nos. 11574092, 61775062,
61378012, 91121023); the National Basic Research Program of China (No.
2013CB921804); the Innovation Project of Graduate School of South China
Normal University.

\section*{APPENDIX}

We consider that the input field is made up of a sequence of pulses with
exactly one photon per pulse, the operator of the input field $\hat{a}%
_{k,in} $ can be expressed as \cite{49}
\begin{equation}
\hat{a}_{k,in}=\int d\omega \xi (\omega )\hat{a}_{k,in}(\omega ),  \tag{A1}
\end{equation}%
where $\xi (\omega )$ is the spectral amplitude for describing the pulse
shape of the single-photon. The operators of the input fields should satisfy
the commutation relation $[\hat{a}_{k,in},\hat{a}_{k,in}^{\dag }]$ $=$ $1$, $%
[\hat{a}_{k,in}(\Omega ),\hat{a}_{k,in}^{\dag }(\omega )]$ $=$ $\delta
(\omega +\Omega )$, and $\int d\omega \left\vert \xi (\omega )\right\vert
^{2}=1$. Generally, the spectrum of the single-photon $S_{k,in}(\omega )$ $%
\equiv $ $\left\vert \xi (\omega )\right\vert ^{2}$ has two forms: the
Gaussian lineshape, or the Lorentzian lineshape, which is in dependence on
its luminescent source.

Now we can define a single-photon state as a superposition of a single
excitation over many frequencies
\begin{equation}
\left\vert 1_{\xi }\right\rangle =\hat{a}_{k,in}^{\dag }\left\vert
0\right\rangle =\int_{-\infty }^{+\infty }d\omega \xi ^{\ast }(\omega )\hat{a%
}_{k,in}^{\dag }(\omega )\left\vert 0\right\rangle ,  \tag{A2}
\end{equation}%
so the correlation functions of the operators of the input fields can be
obtained as
\begin{align}
\left\langle \hat{a}_{k,in}^{\dag }(\Omega )\hat{a}_{k,in}(\omega
)\right\rangle & =\left\vert \xi (\omega )\right\vert ^{2}\delta (\omega
+\Omega ),  \tag{A3} \\
\left\langle \hat{a}_{k,in}(\Omega )\hat{a}_{k,in}^{\dag }(\omega
)\right\rangle & =[\left\vert \xi (\Omega )\right\vert ^{2}+1]\delta (\omega
+\Omega ).  \tag{A4}
\end{align}

\end{document}